\documentstyle[twocolumn,aps,epsfig]{revtex}
\begin{document}
\draft
\title{
Classical Phase Fluctuations in Incommensurate Peierls Chains}

%
\author{Lorenz Bartosch and Peter Kopietz} 
\address{
Institut f\"{u}r Theoretische Physik, Universit\"{a}t G\"{o}ttingen,
Bunsenstrasse 9, 37073 G\"{o}ttingen, Germany}
\date{May 25, 2000}
\maketitle
\begin{abstract}
In the pseudogap regime of one-dimensional incommensurate Peierls
systems, fluctuations of the phase of the order parameter
prohibit the emergence of long-range order and generate a finite
correlation length. For classical 
phase fluctuations, we present exact results for the 
average electronic density of states, the mean localization length,
the electronic specific heat and 
the spin susceptibility at low temperatures. Our results for the
susceptibility give a good fit to experimental data.

\end{abstract}
\narrowtext
%
%
%
Continuous symmetries
in one-dimensional electronic systems
are not spontaneously broken at any finite
temperature $T$. The mean-field prediction
of a finite critical temperature $T_c^{\rm MF}$
is incorrect in this case.
The experimentally observed Peierls transition 
at a finite temperature
$T_c^{\rm 3D} \ll T_c^{\rm MF}$ 
in many quasi one-dimensional conductors
is due to  weak interchain-coupling which
triggers a crossover to three-dimensionality.
In the intermediate temperature regime
$ T_c^{\rm 3D}  \lesssim
T \lesssim T_c^{\rm MF}$,
the physical properties of Peierls chains are
dominated by one-dimensional order parameter fluctuations\cite{Gruener94}.
This is the so-called pseudogap regime where mean-field theory
is not even qualitatively correct.

In this work, we shall present an exact solution of an effective
model for the low temperature thermodynamics of incommensurate Peierls chains.
For incommensurate chains, the order parameter
$\Delta (x)$ is complex, so that
in the pseudogap regime
the generalized Ginzburg-Landau potential  
has the form of a ``Mexican hat'' \cite{Gruener94}. It
is then a good approximation to ignore amplitude fluctuations of 
$\Delta ( x) = | \Delta ( x) | e^{i \vartheta (x )}$ and focus
on the gapless fluctuations of the phase $\vartheta (x)$. 
At long wavelengths and low energies, the thermodynamics 
of phase fluctuations can be described by a classical Hamiltonian which is
formally identical with the kinetic energy of a superfluid \cite{Gruener94},
 \begin{equation}
 H_{\vartheta} = \frac{1}{2} m^{\ast} n_s  \int_0^{L} d x \, v^2_s ( x )
 \; \; , \; \;  v_s ( x) = \frac{ \partial_x \vartheta ( x)}{2 m^{\ast}}
 \; .
 \label{eq:Hphi}
 \end{equation}
Here, $m^{\ast}$ is the effective mass of the electrons, $L$ is the
length of the chain, and 
the one-dimensional density $n_s$ measures the stiffness of the system
with respect to long-wavelength distortions 
of the phase of the order parameter \cite{footnote1}.
A two-dimensional analog of Eq.\ (\ref{eq:Hphi})
has been used by Emery and Kivelson \cite{Emery95}
to explain the pseudogap behavior of underdoped high-temperature
superconductors. In one dimension, the problem is much simpler, because
there is no Kosterlitz-Thouless transition and the thermodynamics
of the phase variable is trivial. 
However, the calculation
of the  electronic properties amounts to solving
a one-dimensional random problem with colored noise.
Usually, problems of this type cannot be solved exactly \cite{vanKampen81}.
At low energies, the electronic degrees of freedom can be described by the
Hamiltonian of the so-called fluctuating gap model (FGM) \cite{Lee73,Lifshits88}
 \begin{equation}
 {H}_{\rm el}  =   
 -  i v_F  \partial_x \sigma_3 +
 \Delta ( x ) \sigma_{+} + \Delta^{\ast} ( x ) \sigma_{-} 
 \label{eq:Hamiltonian}
 \; ,
 \end{equation}
where
$v_F$ is the Fermi velocity, 
$\sigma_{i}$ are the usual Pauli matrices, with
$\sigma_{\pm} = \frac{1}{2}(\sigma_{1} \pm i \sigma_{2})$.
In the pseudogap regime, the gapped amplitude fluctuations are frozen
out, so that we may set
$\Delta (x  ) = \Delta_s e^{ i \vartheta ( x)}$, where
$\Delta_s$ is determined by the local minimum of the generalized
Ginzburg-Landau functional. 
From Eq.\ (\ref{eq:Hphi}),
it is then easy to show that
 \begin{equation}
 \langle \Delta ( x) \Delta^{\ast} ( x^{\prime} ) \rangle =
 \Delta_s^2 \, e^{ -  | x - x^{\prime} | / \xi }
 \; ,
 \label{eq:covdef}
 \end{equation}
where $\langle \ldots \rangle $ denotes the thermodynamic average with
respect to the Hamiltonian $H_{\vartheta}$ given in Eq.\ (\ref{eq:Hphi}),
and the order parameter correlation length is
$\xi = n_s / (2 m^{\ast} T )$.
In this work, we shall calculate the
average electronic density of
states (DOS) $ \rho ( \omega )  = \langle 
{\rm Tr} \delta ( \omega - H_{\rm el}) \rangle$ of the model
defined via Eqs. (\ref{eq:Hphi})--(\ref{eq:covdef})
{\it{exactly for arbitrary  $\xi$}}.

Previously, the DOS of the FGM 
has been calculated assuming a Gaussian distribution of $\Delta (x)$ with
covariance given by 
Eq.\ (\ref{eq:covdef}) \cite{Lee73,Sadovskii79,Tchernyshyov99,Bartosch99b}.
Although in this case the problem is not exactly 
solvable \cite{Tchernyshyov99,Bartosch99b},
a sophisticated algorithm has been 
developed \cite{Sadovskii79}
which produces an expression for $\rho ( \omega)$
which is reasonably close to the exact numerical result for Gaussian
disorder \cite{Bartosch99b}.
However, as explained above,
the assumption of a Gaussian distribution of
$\Delta (x)$ centered at $\Delta = 0$ is rather unphysical in the
pseudogap regime.
It is therefore not surprising that in this regime 
the true behavior of  $\rho ( \omega )$ (to be discussed below) 
is very different from the DOS for
Gaussian disorder. 

The electronic contribution to the thermodynamic
properties of our system can be obtained from the
disorder-averaged free energy 
 \begin{equation}
 F_{\rm el} = - T \int_{- \infty}^{\infty} d \omega \int_0^{L} dx 
 \, \langle
 \rho ( x, \omega
 ) \rangle \ln ( 1 + e^{ - \omega / T })
 \; .
 \label{eq:omegadef}
 \end{equation}
The local DOS  $\rho ( x, \omega )$
can be expressed as 
 $\rho ( x , \omega ) = - {\pi}^{-1}
 {\rm Im} \, {\rm Tr}  \,
 {\cal{G}} ( x , x , \omega + i 0^{+})$,
where the Green function ${\cal{G}} ( x , x^{\prime} , \omega + i 0^{+})$
satisfies
 \begin{equation}
 [ \omega - H_{\rm el} ] \, {\cal{G}} ( x , x^{\prime } , \omega + i 0^{+})
 = \delta ( x - x^{\prime} ) \sigma_0
 \label{eq:Gdef}
 \; .
 \end{equation}
Here, $\sigma_0$ is the $2 \times 2$ unit matrix.
For periodic boundary conditions, the average 
$\langle \rho ( x , \omega ) \rangle$
is independent of $x$ and can be identified
with the average DOS $\rho ( \omega )$. 

Following Ref.\cite{Brazovskii76},
we eliminate the phase of the order parameter
$\Delta ( x ) = \Delta_s e^{ i \vartheta ( x )}$
via a gauge transformation,
 \begin{equation}
 {\cal{G}} ( x , x^{\prime } , \omega)
 =  e^{ \frac{i}{2} \sigma_3 \vartheta ( x) }
 \tilde{\cal{G}} ( x , x^{\prime } , \omega) 
 e^{ - \frac{i}{2} \sigma_3 \vartheta ( x^{\prime}) } 
 \; .
 \label{eq:tildeGdef}
 \end{equation}
The transformed Green function $\tilde{\cal{G}}$ satisfies an equation of the form
(\ref{eq:Gdef}), but with $H_{\rm el}$ replaced by
 \begin{equation}
 \tilde{H}_{\rm el} =
 - i v_F \partial_x \sigma_3 + V ( x ) \sigma_0
 + \Delta_s \sigma_1 
 \; ,
 \label{eq:Heltilde}
 \end{equation}
where 
 $V ( x ) = \frac{v_F}{2} \partial_x \vartheta ( x)$. 
Eq.\ (\ref{eq:tildeGdef}) is a chiral transformation that
eliminates the phase of $\Delta ( x)$ in favor of a
forward scattering random potential $V ( x)$.
The local DOS is invariant under this transformation, so
that
we may replace ${\cal{G}} \rightarrow \tilde{\cal{G}}$
in all expressions involving the DOS. 
The crucial point is now that with
$H_{\vartheta}$ given by Eq.\ (\ref{eq:Hphi}),
the probability distribution
of $V (x)$ is determined by Gaussian white noise, 
with zero average and covariance
$ \langle V ( x ) V ( x^{\prime}) \rangle = 
{v_F^2} (2 \xi)^{-1}  \delta ( x - x^\prime)$.
Due to the Gaussian white noise statistics of $V ( x)$,
the average DOS of our model 
can be calculated exactly in the thermodynamic limit. 
Several methods of 
obtaining the exact $\rho ( \omega )$ are available.
Actually,  Eq.\ (\ref{eq:Heltilde}) is a special case
of the class of random Hamiltonians discussed by 
Hayn and Mertsching \cite{Hayn96}, who calculated the average
DOS by means of the supersymmetry method \cite{Bohr82}.
Alternatively, the DOS can be calculated
within the phase formalism\cite{Lifshits88}. In Ref.\cite{Bartosch00}
a modification of this formalism \cite{Lifshits88,Bartosch99b}
is used to directly obtain 
the integrated
Green function $\Gamma ( \omega)$ defined via 
$ \partial_{\omega} \Gamma ( \omega ) = 
 {\rm Tr}\, \langle {\cal{G}} ( x , x , \omega + i 0^{+}) \rangle$
from the solution of a Fokker-Planck equation.
For $L \rightarrow \infty$, only the stationary solution is needed, 
and we obtain 
  \begin{equation}
 \Gamma ( \omega ) 
  \equiv
 \ell^{-1} ( \omega ) -  i \pi {\cal{N}} ( \omega)
 = \pi \rho_0 \Delta_s
  I_{ - i \nu}^{\prime} ( g ) /  I_{  - i \nu  } (  g ) 
 \; ,
 \label{eq:DOSintres}
 \end{equation}
where $\rho_0 = ( \pi v_F)^{-1}$
is the DOS for $\Delta = 0$, and
$I_{i \nu} ( g )$ is a modified Bessel function with imaginary
index $i \nu$. 
The dimensionless parameters $g$ and $\nu$ are
 \begin{equation}
 g = \frac{4 \Delta_s \xi }{v_F} = \frac{ 2 n_s}{m^{\ast} v_F}
 \frac{\Delta_s}{T}
 \; , \; \; 
 \nu = \frac{ 4 \omega \xi}{v_F} = g \frac{\omega}{\Delta_s}
 \; .
 \label{eq:parameterdef}
 \end{equation} 
Note, that the imaginary
part of $\Gamma ( \omega )$  is proportional to the integrated
average DOS ${\cal{N}} ( \omega )$ which satisfies $\partial_{\omega }
{\cal{N}} ( \omega ) = \rho ( 
\omega)$ while, according to Thouless \cite{Thouless72}, the
real part of $\Gamma ( \omega)$ can be identified with the
inverse mean localization length $\ell^{-1} ( \omega)$,
i.e. the Lyapunov exponent \cite{Lifshits88}.
Using a Wronski relation for $I_{- i\nu} ( g )$ \cite{Abramowitz72}
we get
 \begin{equation}
 {\cal{N}} ( \omega ) = \frac{\rho_0  v_F }{4 \pi \xi }
 \frac{ \sinh  ( \pi \nu )}{
 \left| I_{  i \nu  } 
 (  g ) \right|^2 }
 \; .
 \label{eq:DOSintres2}
 \end{equation}
For the inverse mean localization length we get
\begin{equation}
 \ell^{-1} (\omega) = \frac{\Delta_s}{v_F} \,
 \frac{\partial}{\partial g} \ln | I_{i \nu} ( g) |
 \; .
 \label{eq:locres}
\end{equation}

\begin{figure}
\begin{center}
\epsfxsize8.0cm 
\epsfbox{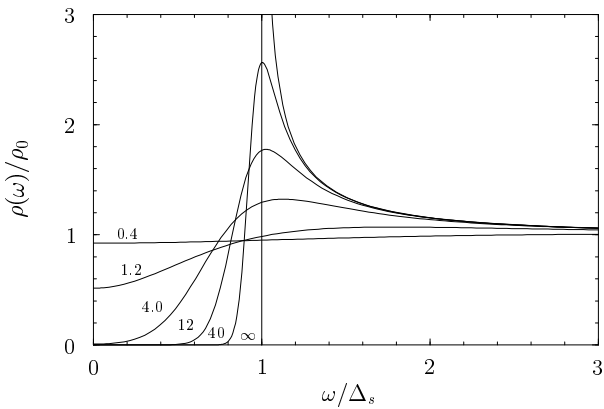}
\vspace{4mm}
\caption{Frequency-dependence of the DOS given in Eq.\ (12) for $g \equiv
  4 \Delta_s \xi /v_F  = 
  0.4,1.2,4.0,12,40$ and $\infty$.} 
\label{fig:dos}
\end{center}
\end{figure}

\begin{figure}
\begin{center}
\epsfxsize8.0cm 
\epsfbox{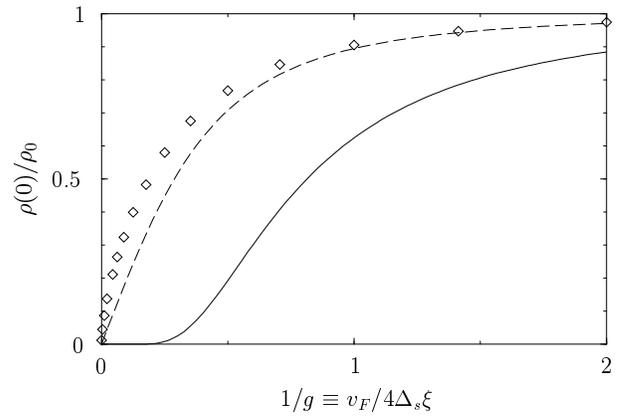}
\vspace{4mm}
\caption{The solid line is a graph of the DOS $\rho(0)$ at the Fermi energy
  for classical phase fluctuations as a function 
  of $1/g \equiv v_F/4 \Delta_s \xi$. For a comparison, the dashed line
  shows the result found in the leading order Born 
  approximation\ [5] and the diamonds give the DOS
  for Gaussian statistics\ [9,13].}
\label{fig:doscomp}
\end{center}
\end{figure}

We now discuss the behavior of the average DOS.
Because $\rho ( \omega )$ is an even function
of $\omega$, we restrict ourselves to  $\omega \geq 0$. 
Using \cite{Abramowitz72}
  $\left| I_{i \nu} ( 0 ) \right|^{2} = (\pi \nu)^{-1} \sinh
   ( \pi \nu ) $
one easily verifies that
${\cal{N}} ( \omega ) \sim \rho_0 \omega$ for $g \rightarrow 0$, so that 
in this limit we recover the result for free electrons with
linearized energy dispersion.  
While for small $g$, the leading corrections can be calculated perturbatively
in powers of $g$, in the pseudogap regime $g \gg 1$,
the behavior of the average DOS is 
quite complicated. It is  convenient
to measure frequencies in units of $\Delta_s$
and to express Eq.\ (\ref{eq:DOSintres2}) in terms of the Bessel function
$J_{ i \nu } ( i g )$ with imaginary index and argument, 
using  $I_{i\nu} ( g ) = e^{ \nu \pi /2 } J_{i \nu } ( i g)$ 
\cite{Abramowitz72}. Defining
$\bar{\omega } = \omega / \Delta_s = \nu / g $, we may write
 \begin{equation}
  \rho ( \omega ) 
 =  \frac{\rho_0}{ 2 \pi g} \frac{ \partial }{\partial \bar{\omega}} 
  \frac{ 1 - e^{ - 2 \pi g \bar{\omega}}}{ 
 \left| J_{ ig \bar{\omega}} ( i g ) \right|^2 }
 \label{eq:Dosres}
 \; .
 \end{equation}
In Fig.\ \ref{fig:dos}, we show a graph of Eq.\ (\ref{eq:Dosres}) for
several values of $g$.
For a more quantitative analysis, we use the 
uniform asymptotic expansion of $J_{ i g \bar{\omega} } ( i g)$ 
 for large $g $ and fixed $\bar{\omega}$
\cite{Abramowitz72} which reveals three different regimes:
First of all, for  $1- \bar{\omega } 
\gtrsim g^{-2/3}$
(i.e. for frequencies sufficiently far below $\Delta_s$),
the average DOS in the pseudogap regime
$g \gg 1$ can be approximated by
\begin{eqnarray}
  \rho ( \omega )  / \rho_0 & \approx &
 2 g ( 1 - \bar{\omega}^2 )^{1/2} \exp [ - 2 g Q ( \bar{\omega} )]
 \nonumber \\
 & & \hspace{+13mm} \times \,
 [ 1 + e^{ - 2 \pi g \bar{\omega}} ]
  \arccos ( \bar{\omega })
 \label{eq:dos1}
 \; ,
\end{eqnarray}
where $Q ( \bar{\omega} ) = ( 1 - \bar{\omega}^2  )^{1/2} - \bar{\omega}
 \arccos ( \bar{\omega})$. 
In particular, for small $\bar{\omega}$, we may expand
$Q ( \bar{\omega}) \approx 1 - \frac{\pi}{2} \bar{\omega} + 
\frac{1}{2} \bar{\omega}^2$, so that
 \begin{equation}
  \rho ( \omega ) / \rho_0 \approx
 2 \pi g e^{-2 g} \cosh[ \pi g  \bar{\omega }]\,
 e^{ - g  {\bar{\omega}^2}  }
 \; , \; g \bar{\omega}^3 \ll 1 
 \; .
 \label{eq:dossmall}
 \end{equation}
Hence, for $\omega = 0$, the DOS is
exponentially small, $\rho ( 0) / \rho_0 \sim 2 \pi g e^{- 2 g}$.
As shown in Fig.\ \ref{fig:doscomp}, 
such a strong suppression of the DOS at the Fermi energy
is a unique feature for classical phase fluctuations, which
is neither reproduced within the
Born approximation \cite{Lee73}
(which predicts $\rho ( 0 ) \propto g^{-1}$),
nor for Gaussian disorder \cite{Bartosch99b} 
(where $\rho ( 0) \propto g^{- \mu}$, with $\mu \approx 0.64$).
The approximation (\ref{eq:dos1}) breaks down
when
$1 - \bar{\omega}$ becomes comparable with 
$g^{-2/3}$. Note that $Q ( 1 - \epsilon ) \sim \frac{2^{3/2}}{3} 
\epsilon^{3/2} $
for  $\epsilon \ll 1$, so that 
$g Q ( \bar{\omega} ) = O ( 1)$
when Eq.\ (\ref{eq:dos1}) ceases to be valid. 
In this case, we have to go back to our exact
result (\ref{eq:Dosres}) which implies 
for $| \bar{\omega } - 1 | 
\lesssim
 g^{-2/3} \ll 1$
 \begin{equation}
   \rho ( \omega )  / \rho_0
 \approx a_1 g^{ {1}/{3}} [ 1 - a_2  g^{{4}/{3}} ( \bar{\omega} - 1 )^2 ]
 \; .
 \label{eq:dos2}
 \end{equation}
Here,
 $a_1 = 2^{- 4/3} \pi^{-1} c_2 / c_1^3 \approx 0.7306$ and
 $a_2 = 2^{2/3} [ 3 (c_2/c_1)^2 - c_1 / c_2 ] \approx 0.3534$,
with $c_1 = {\rm Ai} (0 ) = [3^{2/3} \Gamma ( 2/3)]^{-1}$ and
$c_2 = - {\rm Ai}^{\prime} ( 0) = [3^{1/3}  \Gamma ( 1/3)]^{-1}$,
where ${\rm Ai} ( x )$ is the Airy function.
From Eq.\ (\ref{eq:dos2}), we conclude that, to leading order in $g \gg 1$,
the average DOS 
exhibits a maximum precisely at
$\omega = \Delta_s$, with a hight that 
diverges as  $g^{1/3} \propto \xi^{1/3} \propto T^{-1/3}$ for
$T \rightarrow 0$.
Finally,  for $\bar{\omega } - 1 
 \gtrsim g^{- {2}/{3}}$,
our exact result (\ref{eq:Dosres}) reduces 
to the well known expression for the DOS in the presence of a static
gap,
  $ \rho ( \omega )  / \rho_0
 \approx  \bar{\omega}/\sqrt{\bar{\omega}^2 - 1}$.
At $\bar{\omega} -1 \approx
g^{-2/3}$, this expression smoothly 
matches with the parabola (\ref{eq:dos2}).

\begin{figure}
\begin{center}
\epsfxsize8.0cm
\epsfbox{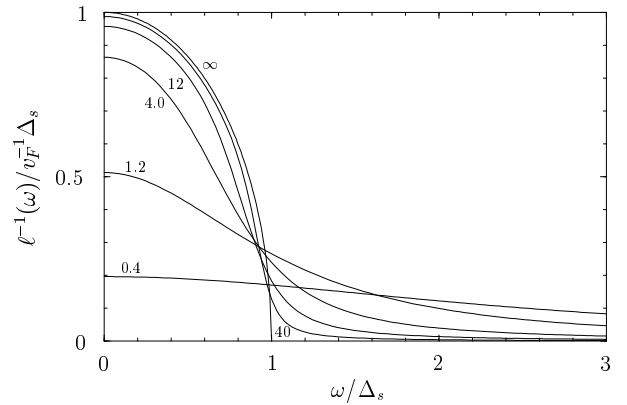}
\vspace{4mm}
\caption{Graph of the inverse localization length for $g \equiv
  4 \Delta_s \xi /v_F  = 
  0.4,1.2,4.0,12,40$ and $\infty$.}                        
\label{fig:loc}
\end{center}
\end{figure}

In Fig.\ \ref{fig:loc}, we show the exact inverse localization
length  $\ell^{-1}(\omega)$ 
given in Eq.\ (\ref{eq:locres}) for several
values of $g$.
For $g \gg 1$ we obtain the following
approximations: 
$\ell^{-1} ( \omega ) \approx (\Delta_s/v_F)
( 1 - \bar{\omega}^2 )^{1/2}$ for 
 $1- \bar{\omega}
 \gtrsim g^{- {2}/{3}}$;
for          
$| \bar{\omega} -1 |
\lesssim g^{- {2}/{3}}$
we find
$\ell^{-1} ( \omega) \approx (\Delta_s/v_F)[
a_3 g^{- {1}/{3}}  - ( 3 g)^{-1} ]$, with
$a_3 = 2^{- {2}/{3}} c_2 /  c_1 \approx 0.4592$;
finally, for
 $\bar{\omega} -1 
 \gtrsim  g^{- {2}/{3}}$
the leading behavior is
$\ell^{-1} ( \omega) \approx (\Delta_s/v_F) [ 2 g (
 \bar{\omega}^2 - 1 ) ]^{-1}$.

Let us now consider the electronic contribution
to the free energy $F_{\rm el}$ defined in Eq.\ (\ref{eq:omegadef}).
For the FGM with a linearized energy dispersion,
Eq.\ (\ref{eq:omegadef})
is ultraviolet divergent, because then the DOS approaches  
a constant for $| \omega| \rightarrow \infty$. 
However, physical quantities involve derivatives of $F_{\rm el}$, which
at low temperatures depend only on 
the low-energy part of the
spectrum and are finite. For convenience, we regularize Eq.\
(\ref{eq:omegadef})
by subtracting from $F_{\rm el}$
the free energy  $F^{\xi = \infty}_{\rm el}$
for an infinite correlation length, where
the gap is static.
After an integration 
by parts,
we express the integral in Eq.\ (\ref{eq:omegadef})
in terms of a fermionic Matsubara sum and obtain
 \begin{equation}
  F_{\rm el} - F_{\rm el}^{\xi = \infty}
 = \frac{2  L \Delta_s   T}{v_F}   \sum_{ n = 0}^{\infty} \left[ 
 \sqrt{ 1 + \bar{\omega}_n^2 }
- \frac{ I^{\prime}_{g \bar{\omega}_n} ( g )}{
 I_{g \bar{\omega}_n} ( g )}
  \right]
 \; ,
 \label{eq:Fres}
 \end{equation}
where $\bar{\omega}_n =  \pi ( 2 n + 1) T / \Delta_s$.
For large $\bar{\omega}_n$,
the term in the square bracket vanishes as $\bar{\omega}_n^{-2}$, so
that the sum converges.
In the pseudogap regime $g \gg 1$,
we may use the 
uniform asymptotic expansion of $I_{g \bar{\omega}_n} ( g)$
for large $g$ \cite{Abramowitz72} to obtain an
expansion of Eq.\ (\ref{eq:Fres}) in powers of $g^{-1} \propto \xi^{-1}
\propto T$.
For $T \ll \Delta_s$ the leading  terms are 
 \begin{equation}  
  F_{\rm el}  - F_{\rm el}^{\xi = \infty} =
 \frac{L}{ 16 \xi}
 \left[ {\Delta_s}  - \frac{v_F}{12 \pi \xi} + O ( \xi^{-2})
 \right]
 \; .
 \label{eq:Omegaxi}
 \end{equation}
The physical interpretation of this result is simple:
Because $\xi$ is roughly the size of domains 
where the order parameter is spatially constant,
the prefactor $L / \xi$ in Eq.\ (\ref{eq:Omegaxi})
can be identified with the number of locally ordered domains
in a system of size $L$. At distances of the order of 
$\xi$, the phase fluctuations  distort the order parameter,
which leads to an increase of the energy. 
In the limit $\Delta_s \xi / v_F \rightarrow \infty $,
the energy scale associated with a twist in the order parameter
is set by $\Delta_s$.  For finite $\xi$ this energy scale decreases,
because the time $\xi / v_F$ it takes for electrons to
propagate over the distance $\xi$ is finite.
This gives rise to the second term in Eq.\ (\ref{eq:Omegaxi}). 
We emphasize that our exact result (\ref{eq:Fres})  gives
the change in the free energy due to phase fluctuations for arbitrary
$\xi$. 

\begin{figure}
\begin{center}
\epsfxsize8.0cm 
\epsfbox{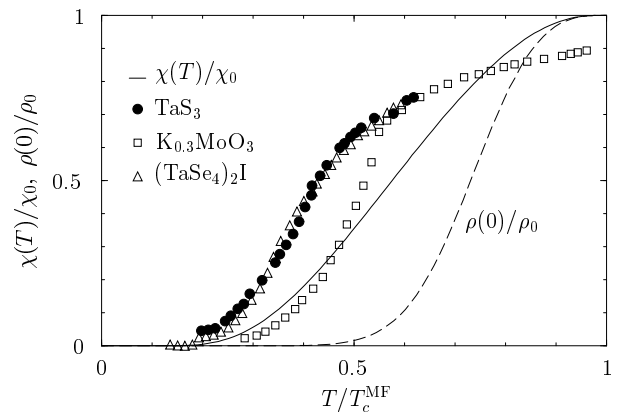}
\vspace{4mm}
\caption{Graph of the susceptibility $\chi(T)$ calculated for $\xi(T) =
  n_s(T)/ 2 m^{\ast} T$ with $n_s(T)$ given in Ref.\ [16] and 
  $\Delta_s(T)$ determined by minimizing a generalized Ginzburg-Landau
  functional. The symbols
represent susceptibility data from Ref.\ [17].
The dashed line is the DOS at the Fermi
  energy.}
\label{fig:spec}
\end{center}
\end{figure}

The low-temperature behavior of the specific heat $C_{\rm el} = - T
\partial^2  F_{\rm 
  el}/ {\partial T^2}$ 
can be calculated analytically.
Keeping in mind that
$\xi = n_s / ( 2 m^{\ast} T )$, we see
that the leading contribution to
$C_{\rm el}$ is due to the
first correction term (involving the energy
$v_F / \xi$) in Eq.\ (\ref{eq:Omegaxi}),
 \begin{equation}
 C_{\rm el}
  \sim 
   ({\pi^2 }/24) ( { n_0 }/{ n_s} )^2  \rho_0 L T
 \; ,
 \label{eq:Cres}
 \end{equation}
where $n_{0} =  m^{\ast} v_F / \pi $, and
we have used the fact that the contribution from
$F_{\rm el}^{\xi = \infty}$
is exponentially small due to the static gap.
Thus, in the pseudogap regime, the electronic specific
heat of Peierls chains is linear in $T$, 
just as the specific heat for non-interacting electrons in one dimension,
$C_{\rm el}^{(0)} \sim \frac{\pi^2}{3} \rho_0 L T$.
Note that $C_{\rm el} / C_{\rm el}^{(0)} = \frac{1}{8} ( n_0 /
n_s)^2$ for $T \rightarrow 0$.
In general, we expect that
$n_s / n_{0}$ is a number of the order of unity for
$T \ll T_c^{\rm MF}$ \cite{footnote2}, so that
$C_{\rm el} / C_{\rm el}^{(0)} = O (1)$. 
In the same regime, we find from Eq.\ (\ref{eq:dossmall})
that $ \rho (0)  / \rho_0 
\sim 4 \frac{n_s }{n_0 } \frac{ \Delta_s}{T} 
\exp [ - \frac{4 }{\pi} \frac{n_s}{n_0} \frac{ \Delta_s}{T} ]$,
i. e. the average DOS at the Fermi energy is exponentially small
(see the dashed line in Fig. \ref{fig:spec}).


Given $ \rho ( \omega )$, 
we may also calculate the spin susceptibility 
$\chi = T^{-1} \int_0^{\infty} d \omega
 \rho ( \omega)  \cosh^{-2} ( \omega / 2 T )$.
A graph of $\chi$ as a function of 
$T / T_c^{\rm MF}$ is shown in Fig. \ref{fig:spec} (solid line).
The low-temperature behavior can again be calculated analytically.
If $g \gg 1$  
but $n_s / n_0 < \frac{1}{4}$, we
find $ \chi \sim  \frac{1}{8}  \rho ( 0 )$
(assuming now $s=2$ for spin degeneracy \cite{footnote1}).
On the other hand, for $n_s / n_0 > \frac{1}{4}$
the frequency integral is dominated by a new saddle point at
$\omega = \cos r$, where $ r = \frac{\pi}{8} \frac{n_0}{n_s}$.
Using Eq.\ (\ref{eq:dos1}), we  obtain 
 \begin{equation}
 \chi / \chi_0  \sim   2 (2 \pi )^{1/2}  r^2 
( \Delta_{r}  / T )^{3/2} \exp [ - \Delta_r /   T ] 
 \; ,
 \label{eq:chires}
 \end{equation}
where  $\Delta_r = \frac{\sin r}{r} \Delta_s$ and
$\chi_0 = 2 \rho_0$ is the susceptibility of free electrons.
$ n_s / n_0 > \frac{1}{4}$ implies
$r < \frac{\pi}{2}$, so that at low temperatures 
the ratio $\chi / 2 \rho ( 0)$
is exponentially large, 
$\chi / 2 \rho ( 0)  \propto 
\exp [ (1 - \sin r ) \Delta_s / ( r T) ]$.
Our graph of $\chi ( T)$ in Fig. \ref{fig:spec}
is quite similar to the corresponding graph given by Lee, Rice, and
Anderson \cite{Lee73}. Note, however, that these authors assumed a
{\it{real}} order parameter and an exponentially large correlation length 
at low temperatures. Because incommensurate
Peierls chains are characterized by a {\it{complex}} order parameter
and a correlation length that diverges only as a power law, 
$\xi \propto T^{-1}$, 
the agreement between the theory of Ref. \cite{Lee73} and
experiments for incommensurate chains \cite{Johnston85} 
seems to be accidental. 
Here, we have shown
that the susceptibility data for
incommensurate Peierls chains can be explained by a non-perturbative
treatment of classical phase fluctuations.
The exponential variation with temperature
is due to the fact that in the pseudogap regime
$- \ln \chi  \propto T^{-1}$. 
Keeping in mind that  our model is strictly
one-dimensional and ignores amplitude fluctuations (which become
important at temperatures of order $T_c^{\rm MF}$), our theoretical
curve for $\chi (T)$ shown in Fig. \ref{fig:spec}
agrees reasonably well with the susceptibility data \cite{Johnston85}.

In summary, we have presented exact results for the
average DOS, the mean localization length, the susceptibility and the
low-temperature thermodynamics of disordered incommensurate Peierls chains
in the pseudogap regime, where only  phase fluctuations
are important. 
In particular, we have derived
the exact frequency-dependence
of $\rho ( \omega)$ which
can be measured by means of
angular integrated photoemission; we predict that
at low temperatures
$\rho ( \omega )$ exhibits a maximum at $\omega = \Delta_s$, 
the hight of which scales as $T^{-1/3}$.

We thank G.\ Gr\"{u}ner and M.\ Grioni for discussions and J.\ Bartosch
for helping us preparing Fig.\ 4.
This work was financially supported by the
DFG (Grants Nos. Ko 1442/3-1 and Ko 1442/4-2).

%

%
%
%


\vspace{-0.4cm}

\vspace{-0.4cm}

\begin{thebibliography}{99}
\bibitem{footnote1}
With the exception of Eq.\ (\ref{eq:chires})
all equations are given for spinless fermions.
Spin degeneracy $s$ can be taken
into account by replacing $n_s \rightarrow s n_s$ and
multiplying all expressions for $F_{\rm el}$  and $C_{\rm el}$
by a factor of $s$.
\bibitem{footnote2} 
Within a gradient  expansion of the generalized Ginzburg-Landau functional,
we find to leading order
$  {n_s}(T)/{n_{0}} =  2 \pi (T/\Delta_s) \sum_{n=0}^{\infty}
(1 + \bar{\omega}_n^2)^{-3/2}$, 
which approaches unity for $T
\rightarrow 0$.
\end{thebibliography}
\end{document}